\documentclass[floats,floatfix,showpacs,amssymb,prd,twocolumn,superscriptaddress,nofootinbib]{revtex4-1}

\usepackage{amsmath,color,graphicx}
\usepackage[normalem]{ulem}
\usepackage{hyperref}
\definecolor{linkcolor}{rgb}{0.0,0.3,0.5}
\hypersetup{colorlinks=true,linkcolor=linkcolor,citecolor=linkcolor,filecolor=linkcolor,urlcolor=linkcolor}
\newcommand\prlsec[1]{\vspace{2mm}\noindent \emph{#1}--}

\def\apj{\rm{ApJ}}
\def\apjl{\rm{ApJ}}

\def\mnras{\rm{MNRAS}}
\def\prd{\rm{PRD}}
\def\prl{\rm{PRL}}
\def\prx{\rm{PRX}}
\def\cqg{\rm{CQG}}
\def\araa{\rm{ARA\&A}}
\def\nat{\rm{Nature}}
\def\aap{\rm{A\&A}}
\def\physrep{\rm{Phys.~Rep.}}

\definecolor{azgreen}{rgb}{0.03,0.47,0.19}
\definecolor{chmagenta}{rgb}{0.54, 0.17, 0.88}

 \newcommand{\Msun}{\ensuremath{M_\odot}}
 \newcommand{\chieff}{\ensuremath{\chi_{\rm eff}}}

\begin{document}

\title{Impact of Bayesian priors on the characterization of binary black hole coalescences}

\author{Salvatore Vitale}
\email{salvatore.vitale@ligo.org}
\affiliation{LIGO, Massachusetts Institute of Technology, 77 Massachusetts Avenue, Cambridge, MA 02139, USA}
\author{Davide Gerosa}
\thanks{Einstein Fellow}
\email{dgerosa@caltech.edu}
\affiliation{TAPIR 350-17, California Institute of Technology, 1200 E California
Boulevard, Pasadena, CA 91125, USA}
\author{Carl-Johan Haster}
\email{haster@cita.utoronto.ca}
\affiliation{Canadian Institute for Theoretical Astrophysics, University of Toronto, 60 St. George Street, Toronto, ON M5S 3H8, Canada}
\author{Katerina Chatziioannou}
\email{kchatziioannou@cita.utoronto.ca}
\affiliation{Canadian Institute for Theoretical Astrophysics, University of Toronto, 60 St. George Street, Toronto, ON M5S 3H8, Canada}
\author{Aaron Zimmerman}
\email{azimmer@cita.utoronto.ca}
\affiliation{Canadian Institute for Theoretical Astrophysics, University of Toronto, 60 St. George Street, Toronto, ON M5S 3H8, Canada}

\date{\today}

\begin{abstract}
In a regime where data are only mildly informative, prior choices can play a significant role in Bayesian statistical inference, potentially 
affecting the inferred physics. 
We show this is indeed the case for some of the parameters inferred from current gravitational-wave measurements of binary black hole coalescences. 
We reanalyze the first detections 
performed by the twin LIGO interferometers using alternative (and astrophysically motivated) prior assumptions. 
We find different prior distributions can introduce deviations in the resulting posteriors that 
impact the physical interpretation of these systems. 
For instance,  (i) limits on the $90\%$ credible interval on the effective black hole spin $\chi_{\rm eff}$ 
are subject to variations of $\sim 10\%$ if a prior with black hole spins mostly aligned to the binary's angular momentum is considered instead of the standard choice of isotropic spin directions, and
(ii) under priors motivated by the initial stellar mass function, we infer tighter constraints on the black hole masses, and in particular, we find no support for any of the inferred masses within the putative mass gap $M \lesssim 5 M_\odot$.
\end{abstract}

\maketitle

\prlsec{Introduction} 
Prior distributions are at the cornerstone of Bayesian statistics, where experimental data is used to update prior knowledge into posterior beliefs about observed phenomena. 
As Bayes himself put it, \textit{``there is reason to expect an event with more or less confidence according to the greater or less number of times in which, under given circumstances, it has happened without failing''} \cite{1763RSPT...53..370B}. 
Prior assumptions inevitably enter into all types of statistical inference, and the prominent role of priors is one of the main advantages of the Bayesian framework: the data analyst is forced to consider and explicitly specify how priors are incorporated into the analysis, thus avoiding the pitfalls of poorly specified assumptions \cite{Jaynes:2003jaq,1995hep.ph...12295D}.

While choosing priors may in some cases be straightforward (for instance, a mass should always be positive), care must be taken when strong knowledge of the expected distribution is not available. 
In particular, in a regime where data are weakly informative and a feeble signal needs to be dug out of instrumental noise, different priors may indeed lead to different 
conclusions.

In this Letter we show that the current statistical inference on some black hole (BH) binary parameters through gravitational-wave (GW) observations does fall in this regime. 
With four likely observations of binary BH coalescences announced, and 
many more expected in the coming years,
GW astronomy is becoming a reality \cite{2016PhRvL.116f1102A,2016PhRvL.116x1103A,2017PhRvL.118v1101A,2016PhRvX...6d1015A}. 
Bayesian analysis has been used to measure the physical parameters of the 
binary coalescences in LIGO's observing runs \cite{2015PhRvD..91d2003V,2016PhRvL.116x1102A}, the BH merger rate~\cite{2016ApJ...833L...1A,2017PhRvL.118v1101A}, as well as to perform tests of general relativity \cite{2016PhRvL.116v1101A,2017PhRvL.118v1101A}.
Some of the physical parameters of the sources, e.g. the total mass (for heavier systems~\cite{2015PhRvD..92b2002G,2015PhRvL.115n1101V,2016MNRAS.457.4499H}) or the chirp mass (for lighter binaries~\cite{1994PhRvD..49.2658C,2017PhRvD..95f4053V,2014PhRvL.112y1101V}), strongly affect the phasing evolution of the GW signal. 
As a consequence, these parameters are usually well measured. 
Others, such as the mass ratio and spins, have a smaller impact on the waveform and are therefore harder to measure. Measurements of these parameters are more directly affected by the chosen prior (see Refs. \cite{FongPrivate,ROSPrivate}).

Accurate statistical inference about BH spins is of crucial astrophysical importance. 
While mass and rates distributions sensibly overlap in many different scenarios (see, e.g., Refs. \cite{2016Natur.534..512B,2016ApJ...824L...8R}),
 spins are arguably the best indicator of BH binary formation channels \cite{2013PhRvD..87j4028G,2017PhRvL.119a1101O,2016ApJ...832L...2R,2016MNRAS.462..844K} (residual eccentricities may also provide a promising avenue; see, e.g., Ref. \cite{2017MNRAS.465.4375N}). 
Several studies have already demonstrated the potential of spin measurements to discriminate between different pathways of BH binary formation and evolution \cite{2017CQGra..34cLT01V,2014PhRvL.112y1101V,2017MNRAS.471.2801S,2017arXiv170408370T,2017arXiv170601385F,2014PhRvD..89l4025G,2017PhRvD..95l4046G,2017ApJ...840L..24F}. 
At the same time, precise and accurate estimation of the individual BH masses can be used to reconstruct their mass function~\cite{2016PhRvX...6d1015A} and to verify if BHs lighter than $5 M_\odot$ exist~\cite{2017MNRAS.465.3254M}.
Quantifying the effect of the prior choice is therefore a crucial step to make solid astrophysical statements using GW data.

\prlsec{Current priors} Parameter estimations for the first GW events GW150914 \cite{2016PhRvL.116f1102A,2016PhRvL.116x1102A,2016PhRvX...6d1014A,2016PhRvX...6d1015A}, GW151226 \cite{2016PhRvL.116x1103A,2016PhRvX...6d1015A}, GW170104 \cite{2017PhRvL.118v1101A} and candidate LVT151012 \cite{2016PhRvD..93l2003A,2016PhRvX...6d1015A} were performed using priors 
(i) uniform in component masses $m_1$ and $m_2$,
(ii) uniform in dimensionless spin magnitude 
$\chi_i = |\mathbf{S}_i|/m_i^2$ 
and (iii) isotropic in spin directions {at the reference GW frequency of 20 Hz}. 
While masses and spin magnitudes are constant up to high post-Newtonian (PN) order, spin directions are subject to change due to orbital plane and spin precession (see, e.g., \cite{1994PhRvD..49.6274A,1995PhRvD..52..821K}). By the time 
BH binaries enter the LIGO band, their spin misalignments may be 
different from those 
at formation, which are needed to discriminate formation channels \cite{2015PhRvD..92f4016G}. 
Fortunately, isotropic spin distributions are kept isotropic when evolved under the 2PN spin precession equations \cite{2004PhRvD..70l4020S,2007ApJ...661L.147B,2015PhRvD..92f4016G}, and the impact of this issue on the usual prior can be neglected (but see $P_5$ below).

Despite being at first sight reasonable, the isotropic spin prior distribution carries important astrophysical consequences. 
It is generally believed that binaries formed in isolation in the galactic field will have, on average, some tendency towards spins aligned with the orbital angular momentum.
On the other hand, spin directions of binaries formed via dynamical interactions in stellar clusters are expected to be isotropically distributed.
We are therefore in a risky situation: we may be biasing our astrophysical inference by assuming \emph{a priori} one of the models we try to discriminate. 

A similar note can be made regarding the choice of distributing spins uniformly in magnitude. 
If we were to base our previous knowledge on other observed BH systems, then moderately high spins should be favored as found in most x-ray binaries \cite{2015PhR...548....1M}. 
On the other hand, core-envelope interactions in 
massive stars may produce BHs with small spins \cite{2002A&A...381..923S,2015ApJ...810..101F} (and it has been suggested that primordial BH spins should also be low \cite{2017arXiv170406573C}). 
One may therefore want to choose a prior which is peaked at either low or high spins, or perhaps even bimodal. 
An agnostic approach would be to consider the BH spins as vectors and draw them uniformly in volume, rather than uniformly in magnitude and isotropic in direction.
Alternatively, one might naively assume black holes form in situations where a random amount of energy goes into the spin, and draw uniformly in specific rotational energy $E_{\rm rot}\equiv 1-\sqrt{1+\sqrt{1-\chi^2}}/\sqrt{2}$.

The spin parameter which is measured best (arguably the only spin parameter which is currently measured at all~\cite{2016PhRvX...6d1015A,2017PhRvL.118v1101A}) is the so-called {effective spin}\footnote{$\chi_{\rm eff}$ is a constant of motion at 2PN \cite{2008PhRvD..78d4021R} and is therefore largely unaffected by the aforementioned issue on the reference frequency. 
} 
\begin{align}	
\chi_{\rm eff} = \frac{\mathbf{S}_1/m_1 + \mathbf{S}_2/m_2}{m_1+m_2}\!\cdot\!\mathbf{\hat L} = \frac{\chi_1 \cos\theta_1 + q\chi_2\cos\theta_2}{1+q},
\label{chieffdefinition}
\end{align}
where $q=m_2/m_1\leq1$ is the mass ratio, and $\theta_i=\arccos({\mathbf{\hat S}_i\cdot\mathbf{\hat L}})$ are the angles between the spins $\mathbf{S}_i$ and the binary's orbital angular momentum $\mathbf{\hat L}$. It is clear from Eq.~\eqref{chieffdefinition} that the mass ratio, spin magnitude and spin direction priors are all entangled in determining the prior distribution of $\chi_{\rm eff}$. %

\prlsec{Prior choices}
In order to explore some of these issues and gauge the impact of priors on Bayesian inference, we have reanalyzed the BH coalescences 
detected by LIGO 
during its first observing run (O1)
using a variety of alternative prior distributions.
Results have been obtained using the nested sampling algorithm implemented in \textsc{LALInference}~\cite{2015PhRvD..91d2003V} and a reduced-order quadrature \cite{2016PhRvD..94d4031S} 
implementation of the \textsc{IMRPhenomPv2} waveform model, which partially accounts for spin precession effects through a single parameter $\chi_{\rm p}$~\cite{2014PhRvL.113o1101H}. 
We restrict our study to (detector-frame~\cite{2017PhRvD..95f4052V}) chirp masses  $8 M_\odot \leq M_c \leq 45 M_\odot$, mass ratios $q\geq 1/8$ and dimensionless spin magnitudes $\chi_i\leq 0.89$ \cite{2016PhRvD..94d4031S}. 
These restrictions are not a problem for our study since for none of the runs we performed the posterior distributions had support near these boundaries. 
We have 142 analyzed the 32-sec data frames publicly released by the 143 LIGO/Virgo Collaboration (\href{https://losc.ligo.org}{\texttt{losc.ligo.org}}),
 using the \texttt{BayesWave} algorithm \cite{2015PhRvD..91h4034L,2015CQGra..32m5012C} to estimate the on-source power spectral density needed for the likelihood evaluations~\cite{2015PhRvD..91d2003V},
marginalizing over calibration uncertainties as in Refs.~\cite{2016PhRvX...6d1015A,2017PhRvL.118v1101A}.

Each event was analyzed multiple times, using one of the following priors: %
uniform in individual masses 
and spin magnitudes, 
isotropic in spin direction 
$(P_1$, the default choice used in LIGO analyses\footnote{This is essentially the same prior used in \cite{2016PhRvL.116f1102A,2016PhRvL.116x1102A,2016PhRvX...6d1014A,2016PhRvL.116x1103A,2017PhRvL.118v1101A,2016PhRvX...6d1015A} apart from the $M_c$, $q$ and $\chi_i$ limitations required by the ROQ implementation.});
uniform in individual masses 
and rotational energy of the BHs,
isotropic in spin direction $(P_2)$; 
uniform in individual masses, 
spin vectors uniform in volume 
$(P_3$, volumetric); 
uniform in individual masses, 
bimodal in spin magnitudes $P(\chi_i)\propto \exp [-(\chi_i-\mu_1)^2/(2\sigma^2)]+\exp[-(\chi_i-\mu_2)^2/(2\sigma^2)]$ with $\mu_1=0,\mu_2=0.89,\sigma=0.1$, 
isotropic in spin directions 
$(P_4)$ \cite{2017arXiv170200885Z,2017ApJ...842..111H,2017arXiv170607053B}; 
uniform in individual masses 
and spin magnitude,
peaked around alignment for spin direction $P(\theta_i)\propto \exp[-(\cos\theta_i-\mu)^2/2\sigma^2]$ where $\mu=1, \sigma=1-\cos(10^\circ)$
 $(P_5$, c.f.~Ref.~\cite{2017arXiv170408370T}, which suggested these parameters could be inferred with $\mathcal{O}(50)$ observations); 
 power law in primary's mass $P(m_1)\propto m^{\alpha}$ with $\alpha=-2.3$ 
 (as in Kroupa's initial mass function for massive stars~\cite{2001MNRAS.322..231K}, c.f.~also Ref.~\cite{2010ARA&A..48..339B} and references therein),
uniform in secondary's mass,
uniform in spin magnitude, 
isotropic in spin direction 
$(P_6)$; 
power law in $m_1$ with $\alpha=-2.3$, 
{logistic} prior in the mass ratio $P(q)\propto 1/\{1+\exp[-k(q-q_0)]\}$ with $k=20, q_0=0.8$ (this is meant to mimic numerical results of BH mergers in globular clusters~\cite{2016PhRvD..93h4029R}), 
uniform in spin magnitude,
isotropic in spin direction 
$(P_7)$;
uniform in individual masses, 
 Gaussian around zero for dimensionless spin magnitude $P(\chi_i)\propto \exp[-(\chi_i-\mu)^2/(2\sigma^2)]$ with $\mu=0,\sigma=0.1$ 
$(P_8)$.
In what follows we occasionally refer to the evidence of the priors. What is meant is the evidence of a Bayesian \emph{model} where the spins and masses are distributed as in that prior.

\begin{table}
\begin{centering}
\begin{tabular}{c ccc ccc ccc}
\hline
\hline
\noalign{\smallskip}
$\quad$&\multicolumn{3}{c}{GW150914} & \multicolumn{3}{c}{GW151226} & \multicolumn{3}{c}{LVT151012} \\ 
& $D^{\chi_{\rm eff}}_{\rm KL}$ & $D^{\chi_{\rm p}}_{\rm KL}$ & $\log_{\!10}\!\mathcal{O}$& $D^{\chi_{\rm eff}}_{\rm KL}$ & $D^{\chi_{\rm p}}_{\rm KL}$ & $\log_{\!10}\!\mathcal{O}$ & $D^{\chi_{\rm eff}}_{\rm KL}$ & $D^{\chi_{\rm p}}_{\rm KL}$ & $\log_{\!10}\!\mathcal{O}$ \\
\hline
\noalign{\smallskip}
$P_1$ & 1.02 & 0.03 & ---     & 1.93 & 0.21 & ---     & 0.53 & 0.03 & ---   \\
$P_2$ & 1.36 & 0.06 & -0.3  & 1.78 & 0.04 & 0.0   & 0.89 & 0.05 & -0.1   \\
$P_3$ & 1.52 & 0.09 & -0.4  & 1.76 & 0.02 & 0.0   & 0.95 & 0.04 & 0.0  \\
$P_4$ & 0.88 & 0.12 & 0.0   & 2.56 & 0.70 & -0.1  & 0.61 & 0.12 & -0.1   \\
$P_5$ & 4.21 & 1.75 & -1.7 & 0.82 & 0.21 & 0.0 & 0.22 & 0.07 & 0.5 \\
$P_6$ & 0.96 & 0.01 &0.1 & 2.12 & 0.08 & 0.4  & 0.24 & 0.00 & 0.4   \\
$P_7$ & 0.93 & 0.06 & 0.4  & 2.63 & 0.02 & 0.4   & 0.26 & 0.01 & 0.5 \\
$P_8$ & 0.14 & 0.07 & 0.3  & 4.82 & 0.70 & -1.7   & 0.03 & 0.02 & -0.1  \\
\noalign{\smallskip}
\hline
\hline
\end{tabular}
\end{centering}
\caption{For each of the three O1 events, we show the KL divergence $D_{\rm KL}$ on (i) $\chi_{\rm eff}$ and (ii) $\chi_{\rm p}$ measuring the information gain between prior to posterior in bits, and (iii) the Bayesian odds ratio $\log_{10}\mathcal{O}$ of each single analysis $P_n$ compared to the standard one $P_1$.   $P_n$ ($P_1$) is preferred if $\log_{10}\mathcal{O}$ is positive (negative).  For comparison, the KL divergence between the $\chi_{\rm eff}$ $P_1$ prior and a uniform distribution over the same range is $0.82$ bits. The log odds have an uncertainty of  $\pm 0.04$.}
\label{dklodds}
\end{table}

\begin{figure*}
\begin{center}
\includegraphics[width=\textwidth]{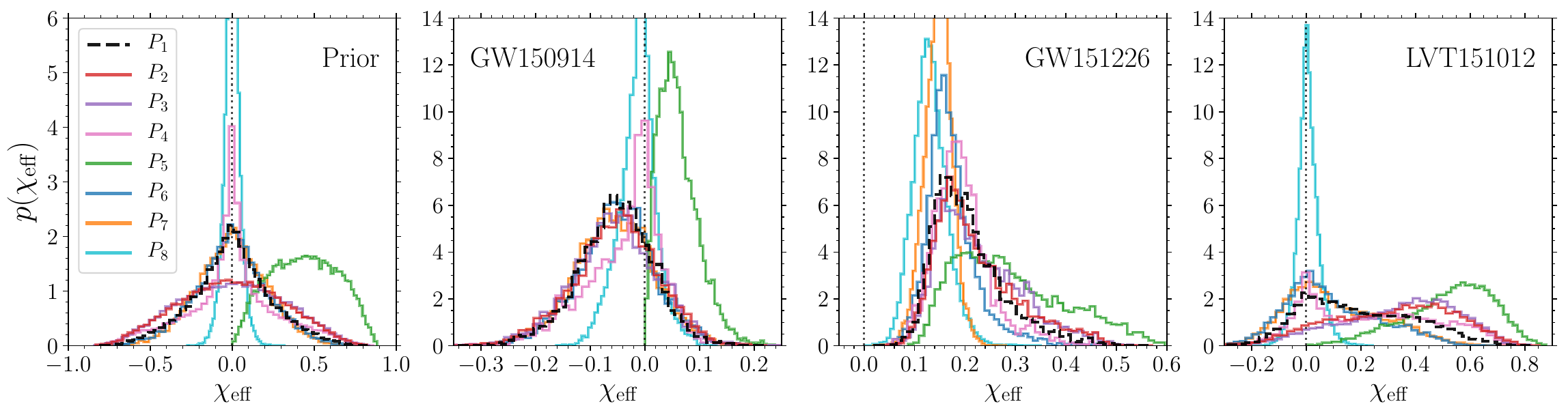}
\caption{\label{fig:chieffposterior} Marginalized prior and posterior distributions on $\chi_{\rm eff}$  for each of the current GW events.
Dashed black line shows results obtained with canonical prior choice $P_1$ (uniform in masses, spin magnitude and isotropic in spin directions), while lighter colored lines shows our  alternative prior assumptions $P_2 -  P_8$. 
In particular, the event GW150914 is compatible with nonspinning BHs, while $\chi_{\rm eff}<0$ is excluded at the 99\% credible level for GW151226 for all prior choices. 
}
\end{center}
\end{figure*}

\prlsec{Spins }
Marginalized prior and posterior distributions in $\chi_{\rm eff}$ are shown 
in Fig.~\ref{fig:chieffposterior}.
Table~\ref{dklodds} shows the corresponding values of the Kullback-Leibler (KL) divergence~\cite{kullback1951} (a measurement of the information gain between prior and posterior, or equivalently the relative entropy between two distributions)
for both spin parameters $\chi_{\rm eff}$ and $\chi_{\rm p}$, together with the 
odds ratio $\log_{10} \mathcal{O}$ between the posteriors obtained with each of our choice $P_n$ and the default analysis $P_1$.

For GW150914, all results are compatible with zero $\chi_{\rm eff}$, which might come from a combination of small spins or in-plane spins, see Eq.~\eqref{chieffdefinition}. Small spin magnitudes are preferred for some prior choices.
The bimodal prior $P_4$ is particularly interesting: when restricted to choose between high and low spins only, the data clearly favor the low spin mode for both objects, with $p(\chi_1<0.3)=0.78, \: p(\chi_2<0.3)=0.68$.  If aligned spins are assumed ($P_5$), the posterior distribution still peaks close to the nonspinning configuration $\chi_{\rm eff}=0$ with high information gain ($D_{\rm KL}^{\chi_{\rm eff}}\sim 4.2$, $D_{\rm KL}^{\chi_{\rm p}}\sim 1.7$) but low odds ($\mathcal{O}\sim 0.02$), thus suggesting the algorithm cannot 
model the data equally well if both tilts are low. 
 Together, these observations indicate that the conclusion that $\chi_i \lesssim 0.3$ for GW150914 is robust to changes in the prior.

Conversely, for GW151226 all priors exclude $\chi_{\rm eff}\leq0$ at the 99\% credible level,
thus confirming  with extremely high significance that at least one of the two BHs was spinning~\cite{2016PhRvL.116x1103A}.
 The bimodal prior $P_4$ in this case favors the high spin mode for the spin of the heavier BH, while both modes are equally likely for the less massive object: $p(\chi_1>0.445)=0.83; \: p(\chi_2>0.445)=0.59$
[but $p(\chi_1  < 0.445 \cup \chi_2 < 0.445) < 0.01$].
The case of the aligned-spin prior $P_5$ presents important astrophysical consequences. 
With odds very similar to the $P_1$ run, $P_5$ allows for posterior values of $\chi_{\rm eff}$ as large as $0.49$ at the 90\% credible level, thus allowing for moderately large spin magnitudes. 
Interestingly, priors $P_6$ and $P_7$ lead to narrower posteriors compared to $P_1$, as they both place more support in the $q\lesssim 1$ region, thus partly breaking the $q$-$\chi_{\rm eff}$ degeneracy~\cite{2016PhRvX...6d1015A}.  This effect is less important for GW150914 because of its higher total mass. 

Finally, prior effects are even more pronounced for LVT151012. 
This is not surprising, as its lower signal-to-noise ratio indicates the data are less informative. In particular, two modes appear to be present in the marginalized posterior of the effective spin, located at $\chi_{\rm eff}\sim 0$ and $\chi_{\rm eff}\sim 0.5$ respectively. 
Which of the two modes is preferred depends on the prior distribution: $P_4$, $P_6$, $P_7$, and $P_8$ prefer the low-$\chi_{\rm eff}$ mode, while $P_2$, $P_3$ and especially $P_5$ favor higher values of $\chi_{\rm eff}$. 

As shown in Table~\ref{dklodds}, the values of $D_{\rm KL}^{\chi_{\rm p}}$ are close to zero for most priors and all events, thus indicating the  $\chi_{\rm p}$ prior distribution is typically returned as a posterior  almost unchanged by the data. As already mentioned, an exception is $P_5$ for
 GW150914,
 which however is disfavored at $\mathcal{O}\sim 0.02$. 
For GW151226, the $P_4$ prior yields a posterior on $\chi_{\rm p}$  which is quite different from the prior, $D_{\rm KL}^{\chi_{\rm p}}=0.7$ bits,  
and models that data comparably well.
This happens because the prior of the spins magnitude is bimodal, while the posterior prefer the high spin mode for the primary.

Since values of both the (detector-frame) masses and of $\chi_{\rm eff}$ are similar, we expect the most recent event GW170104 to present the same trends as GW150914 (perhaps with broader posterior distributions due to the lower signal-to-noise ratio). 
We verified this by constructing a software replica 
of  GW170104 with parameters consistent to those presented in Ref.~\cite{2017PhRvL.118v1101A} and simulating its noise power spectral density using the same method as Ref.~\cite{2016LRR....19....1A}. 
Variations in the 90\% credible interval to $\chi_{\rm eff}$ of up to $\sim 30\%$ are observed: while priors which include both high and misaligned spins all return 90\% credible intervals of $-0.4 \lesssim \chi_{\rm eff} \lesssim 0.1$, inference with the low-spin prior $P_8$  returns $-0.07 < \chi_{\rm eff}< 0.05$, while  $0.03 < \chi_{\rm eff}< 0.23$ if aligned spins are assumed  ($P_5$). 

\begin{figure}
\begin{center}
\centering
\includegraphics[width=0.85\columnwidth,clip=true]{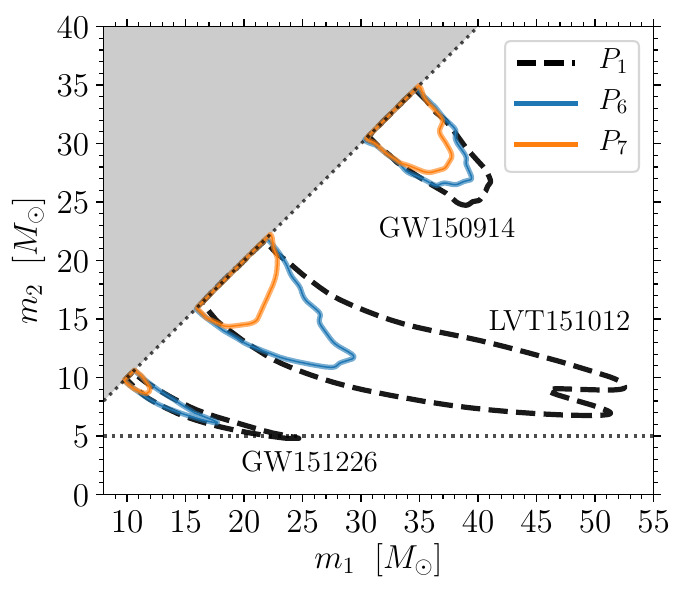}

\caption{\label{m1m2} 
Posterior distribution for the individual masses $m_1$ and $m2$ (with $m_2<m_1$) derived using the canonical prior $P_1$ and two other choices motivated by stellar physics  ($P_6$ and $P_7$). For GW151226, the region $m_2<5 M_\odot$ is excluded at $>99$\% probability for both $P_6$ and $P_7$. 
}
\end{center}
\end{figure}

\prlsec{Component masses}
Different choices for the component mass priors also carry important astrophysical implications. 
Figure~\ref{m1m2} shows posterior distributions of the two BH masses for the 
three O1
 events using the default $P_1$ prior,\footnote{We notice that the $P_1$ posteriors on LVT151012 are  wider than those presented in \cite{2016PhRvX...6d1015A}. 
This is due to differences in the way the power spectral density was estimated compared to this analysis.} as well as $P_6$ and $P_7$. 
Both $P_6$ and $P_7$ have larger prior support for binary mass ratio close to unity compared to $P_1$.
This additional weight at comparable masses has a visible effect on the posteriors.
While presenting odds similar to $P_1$, the resulting posterior distributions for $P_6$ and $P_7$ now prefer the region closer to the $m_1=m_2$ line in Fig.~\ref{m1m2}. 
In particular, when GW151226 is analyzed with any of these two priors, the 99\% credible interval for the source-frame mass of the secondary object is above 5\Msun.

This point has serious consequences in the astrophysical interpretation of the results.
Electromagnetic observations of neutron stars and stellar-mass BHs (c.f. Ref.~\cite{2015PhR...548....1M} and references therein) hint at a putative mass gap between highest neutron star masses ($m \lesssim 3 M_\odot$) and the lowest BH masses ($m\gtrsim 5 M_\odot$). 
Current measurements, however, are not conclusive since the lack of BHs at lower masses could be entirely due to selection effects \cite{2012ApJ...757...36K}. 
The confirmation or exclusion of the mass gap by GW observations is expected to provide unique insights on stellar collapse and compact-object formation \cite{2002ApJ...572..407B}.
Figure~\ref{m1m2} shows that, when analyzed with priors motivated by stellar physics like Kroupa's initial mass function, GW data for GW151226 
are fully consistent with
 the existence of a mass gap. 
Both $P_6$ and $P_7$ are slightly favored over the default prior, with $\mathcal{O} \approx 2.5$.
A careful considerations of priors may thus be important to securely discriminate between BHs and neutron stars \cite{2013ApJ...766L..14H,2015ApJ...798L..17C,2014PhRvD..89j4023C,2015MNRAS.450L..85M,2015ApJ...807L..24L,2016ApJ...825..116F} and thus establish the presence of the mass gap between the two classes of sources.  

Finally, we have verified that the marginalized chirp mass posteriors are stable over the change of priors. For GW150914, all cases except $P_5$ (which, however, presents lower odds) yield posterior medians within a $\sim0.5 M_\odot$ interval. The median for $P_5$ is $1M_\odot$ larger than that of $P_1$ (for comparison the 90\% credible interval for the $P_1$ run is $\sim 3M_\odot$). For GW151226, all runs yield posterior medians within a $\sim 0.03M_\odot$ interval, compared with a 90\% credible interval of $\sim 0.6 M_\odot$ for the $P_1$ run.

\prlsec{Conclusions and future prospects} 
In this Letter we have shown how different prior choices can influence the statistical inference on the parameters of binary BHs. 
We have estimated the parameters of the first GW events detected 
under different prior assumptions and verified that both the 
component masses
and the effective spin are impacted by the choice of the prior.
For example, %
for GW150914  
and the software replica of GW170104
the effective spin can be made close to zero using a prior that prefers small spin magnitudes, without significant loss of evidence. Conversely, for GW151226 all priors we used give a posterior density for \chieff{}  which is positive at the $99\%$ credible interval.
Using the default priors for GW151226 results in a source-frame mass posterior for the lighter BH which has support in the suggested mass gap between BHs and neutron stars.
We have shown that this conclusion does not hold if a different,
yet reasonable, prior on the masses (i.e. a power law
modeled on the initial stellar mass function with and without a logistic
function on the mass ratio) is used.

As exemplified by the case of LVT151012, the effect of different prior choices is more severe for weak GW signals. 
Data will be more informative for future loud events and, eventually, more and more physical conclusions will become robust with respect to the details of the prior choice. 
This point will be specifically addressed in future work, together with a wider variety of prior distributions more carefully modeled on BH formation pathways, along with the effect of the \mbox{PN spin evolution \cite{inprep}.} Other investigations of prior effects in GW data analysis are also underway \cite{2017arXiv170903095W}.

We hope our work may spark  new efforts at incorporating a range of prior choices into inferences about the progenitors of binary BHs. Instead of attempting to explain constraints obtained from a single prior choice, we argue different astrophysical models should be used as priors in the data analysis process. Model selection should then be applied to assess which model better explains the observations. 
For instance, posteriors calculated under different priors naturally enter in hierarchical analyses~\cite{2010ApJ...725.2166H,2017arXiv170601385F,2017arXiv170408370T,2017arXiv170901943W,2017ApJ...840L..24F,	
2017ApJ...846...82Z}, or when one wants to calculate the branching ratio between populations~\cite{2017CQGra..34cLT01V}. More simply, one can just use the evidences calculated for various priors, which we report in Table~\ref{dklodds}, to calculate the odds ratios between models. In this scenario, a cumulative odds given multiple detections can be easily produced (e.g. Ref.~\cite{2014PhRvD..89h2001A}).

Ultimately, statistical inference consists of the continual update of one's current (prior) knowledge in the face of new observational evidence. 
The dependence of one's beliefs on prior knowledge should be viewed as a strength rather than a weakness as we approach new observations, since priors allow us to account for all past evidence on the subject.

\vspace{0.2cm}
\prlsec{Acknowledgments}
We thank Thomas Dent, Jim Fuller, Chris Pankow, Harald Pfeiffer, Vivien Raymond and Leo Stein for various stimulating discussions. We thank Christopher Berry, James Clark, Will Farr, Heather Fong, Prayush Kumar, Richard O'Shaughnessy, Harald Pfeiffer, and John Veitch for discussions on the use of the volumetric prior and making their implementation public.
We thank Lisa Barsotti for providing a simulated noise spectral density for the analysis of the GW170104 software replica. 
S.V.~acknowledges support of MIT through the Solomon Buchsbaum Research Fund, the National Science Foundation, and the LIGO Laboratory. 
LIGO was constructed by the California Institute of Technology and Massachusetts Institute of Technology with funding from the National Science Foundation and operates under cooperative agreement PHY-0757058.
D.G.~is supported by NASA through Einstein Postdoctoral Fellowship Grant No. PF6-170152 awarded by the Chandra X-ray Center, which is operated by the Smithsonian Astrophysical Observatory for NASA under Contract NAS8-03060. 
The authors %
acknowledge the LIGO Data Grid clusters.
This project started during the workshop {``Strong Gravity and Binary Dynamics with Gravitational Wave Observations''}, funded through NSF CAREER Grant No. PHY-1055103. 
This paper carries 
LIGO Document Number P1700176.

\bibliography{priordraft}

\end{document}